\documentclass[reprint,amsmath,amssymb, aps,pra]{revtex4-2}

\usepackage{graphicx}
\usepackage{dcolumn}
\usepackage{bm}

\begin{document}
\title{Observation of Subnatural-Linewidth Biphotons In a Two-Level Atomic Ensemble}
\author{Jyun-Kai Lin$^{1}$}
\email{jyunkai@mx.nthu.edu.tw}
\author{Tzu-Hsiang Chien$^{1}$}
\author{Chin-Te Wu$^{1}$}
\author{Ravikumar Chinnarasu$^{1}$}
\author{Shengwang Du$^{2,3,4,5}$}
\author{Ite A. Yu$^{1}$}
\author{Chih-Sung Chuu$^{1}$}
\email{cschuu@phys.nthu.edu.tw}

\affiliation{$^{1}$Department of Physics and Center for Quantum Science and Technology,
National Tsing Hua University, Hsinchu 30013, Taiwan}
\affiliation{$^{2}$Elmore Family School of Electrical and Computer Engineering, Purdue University, West Lafayette, Indiana 47907,USA}
\affiliation{$^{3}$Department of Physics and Astronomy, Purdue University, West Lafayette, Indiana 47907,USA}
\affiliation{$^{4}$Purdue Quantum Science and Engineering Institute, Purdue University, West Lafayette, Indiana 47907,USA}
\affiliation{$^{5}$Department of Physics, The University of Texas at Dallas, Richardson, Texas 75080, USA}

\begin{abstract}
Biphotons and single photons with narrow bandwidths and long coherence times are essential to the realization of long-distance quantum communication (LDQC) and linear optical quantum computing (LOQC).In this Letter, we manipulate the biphoton wave functions of the spontaneous four-wave mixing in a two-level atomic ensemble with a single-laser pump scheme. Our innovative experimental approach enables the generation of biphotons with a sub-MHz bandwidth of 0.36 MHz, a record spectral brightness of $2.28\times10^7$~${\rm s}^{-1}{\rm mW}^{-1}{\rm MHz}^{-1}$, and a temporally symmetric wave packet at moderate optical depth. The strong non-classical cross-correlation of the biphotons also enables the observation of heralded sub-MHz-linewidth single photons with a pronounced single-photon nature. The generation of sub-MHz-linewidth biphotons and single photons with a two-level atomic ensembles not only finds applications in quantum repeaters and large cluster states for LDQC and LOQC but also opens up the opportunity to miniaturize the biphoton or single-photon sources for chip-scale quantum technologies. 
 
\end{abstract}


\maketitle

Biphotons and single photons play an essential role in photonic quantum technologies \citep{1}. Conventionally, spontaneous parametric down-conversion (SPDC) in a nonlinear crystal has been an admirable candidate for generating biphotons or heralded single photons for more than three decades. However, the constraint of the phase-matching condition on the SPDC-based biphotons causes the bandwidths to be too broad to efficiently interact with the atoms, thus limiting their applications. For example, long-distance quantum communication (LDQC)~\citep{2}, linear optical quantum computing (LOQC) \citep{3}, and other applications involving quantum repeaters require narrowband and temporally long biphotons or single photons. The storage efficiency of the electromagnetically-induced-transparency-based (EIT-based) quantum memories also favors biphotons of narrow bandwidths and long temporal lengths~\citep{4,5,6,7,8,9,10,11,12,13,14}. In addition, the short coherence time of SPDC-based biphotons or single photons also makes it extremely difficult to observe the interference between independent photon sources \citep{15}.Therefore, a challenging obstacle occurs if we are dedicated to develop quantum technologies with only SPDC-based biphotons in our hands.

Today, spontaneous four-wave mixing (sFWM)~\cite{20,21,Du:08} provides an astonishing tool to generate narrowband biphotons. With the help of EIT~\cite{16}, the biphoton bandwidth can be further reduced from the natural linewidth of the atoms to the subnatural linewidth level, offering a promising solution to build state-of-the-art quantum technologies. Manipulating and shaping the wave packets of these narrowband and temporally long biphotons \cite{22,23,24,25} are also feasible, making possible high-efficiency quantum memory \cite{26}, electro-optic modulation of single photons \cite{27}, quantum key distribution with high key creation efficiency \cite{28,29, Liu:13,Kao2023}, controlled single-photon absorption and re-emission \cite{30,Wu2017}, purification of single and entangled photons~\cite{Feng2017,31,Chuu2021}, and quantum computing with a single photon in high dimensions \cite{Wen2024}.

In typical experiments utilizing sFWM in Rb atoms, hyperfine structures of the D1 ($5^{2}S_{1/2}\rightarrow5^{2}P_{1/2}$) and D2 ($5^{2}S_{1/2}\rightarrow5^{2}P_{3/2}$) transitions are employed as pump and coupling fields to obtain large group delays. Thus, the EIT-based biphoton generation is usually
viewed as a four-level double-$\Lambda$ atomic system. Apart from the EIT regime, the paired photons from sFWM can also be generated in the waveform of damped Rabi oscillation when the optical depth (OD) is low \citep{20, Du:08}. In this regime, the biphoton bandwidth is much broader than that of the EIT-based biphotons. However, by manipulating the multimode biphoton wave function and controlling the quantum interference of two possible sFWMs, subnatural-linewidth biphotons can be efficiently generated at low OD in the damped Rabi oscillation regime~\citep{32}. Although the sFWM in the four-level double-$\Lambda$ atomic system can generate subnatural-linewidth biphotons, the miniaturization of these biphotons sources for the chip-scale quantum device will be challenging and complex due to the need of multiple laser systems \citep{33,34}. Since the realization of sophisticated quantum technologies requires an integrated optics architecture for improved performance and scalability, the development of miniature atom-based instruments is indispensable. 

The possibility of sFWM-based biphoton generation in a two-level atomic system has also been explored theoretically and experimentally~\citep{35,36,37,38}. Compared with a four-level system, a single laser system is sufficient for realizing the sFWM in a two-level system. Moreover, because the biphoton bandwidth is only limited by the dephasing rate of the inhomogeneously broadened ground state, the generation of sub-MHz-linewidth biphotons is also feasible. The sFWM in a two-level system is thus a promising compact source of sub-natural-linewitdh biphotons and heralded single photons. Nevertheless, the biphoton generation in a two-level system has never been achieved in the subnatural- or sub-MHz- linewidth regime. The generation of heralded single photons in a two-level system has also never been successful because of the weak nonclassical correlation of the biphotons. 

In this Letter, we demonstrate the generation of subnatural-linewidth biphotons in a two-level atomic ensemble. Our innovative experimental approach enables the generation of biphotons with a bandwidth of 0.36 MHz and a record spectral brightness of $2.28\times10^7$~${\rm s}^{-1}{\rm mW}^{-1}{\rm MHz}^{-1}$ at moderate optical depth. The strong non-classical cross-correlation of the biphotons also enables the observation of heralded sub-MHz-linewidth single photons that exhibit antibunching in the Hanbury-Brown and Twiss interferometer. Moreover, the single photons have a temporally symmetric double-exponential wave packet, which is advantageous for efficient quantum storage and state transfer between quantum nodes~\cite{Cirac1997}. These achievements are made possible with three key steps: (1) the decoherence of sFWM in the two-level system is suppressed, (2) the uncorrelated photon pairs and the elastic scattering of the pump field are reduced, and (3) the biphoton wave function is spectrally manipulated to generate single-mode biphotons. Our work demonstrates a novel way of achieving sub-natural and sub-MHz linewidth for biphotons and single photons with a two-level atomic ensemble.

\begin{figure}
\includegraphics[width=8cm,height=10cm]{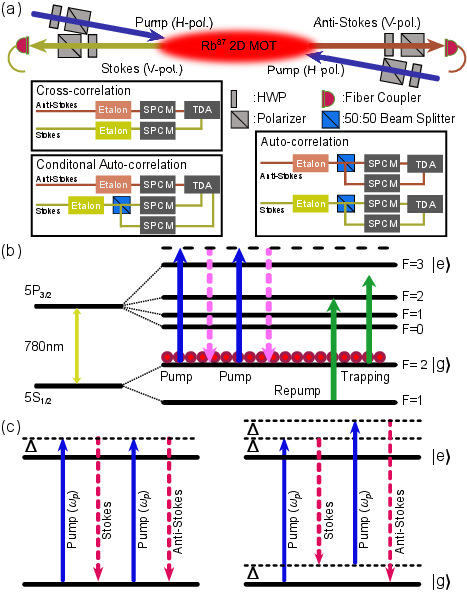}
\caption{\label{fig:2}(a) Experimental setup for generating narrowband biphotons and single photons in a two-level atomic ensemble. SPCM: single-photon counting module, TDA: time-to-digital analyzer, HWP: half wave plate. (b) Energy-level diagram for generating sub-MHz-linewidth biphotons and single photons. (c) In the presence of counter-propagating pump fields at frequency $\omega_{p}$, the Stokes and anti-Stokes photons are generated in two possible sFWM processes: the two-photon resonance (left panel) and three-photon resonance (right panel).}
\end{figure}

The schematic of our experimental setup is illustrated in Fig.~\ref{fig:2}(a). Biphotons are generated using an elongated cloud of $^{87}$Rb atoms in a standard six-beam two-dimensional magneto-optical trap (2D MOT) where the relevant atomic levels are $\left|g\right\rangle =\left|5S_{1/2},F=2\right\rangle $ and $\left|e\right\rangle =\left|5P_{3/2},F=3\right\rangle $. The atomic cloud has a length of about 5 mm and the OD is 12. Two counter-propagating pump fields are aligned at an angle of 4° with respect to the longitudinal \textit{z}-axis and the two identical horizontally polarized pump fields $\omega_{p}$ are blue-detuned from the cycling transition with a detuning $\Delta$. Through sFWM, phase-matched Stokes ($\omega_{s}$) and anti-Stokes ($\omega_{as}$) photons are spontaneously generated in the backward-wave configuration. A set of polarization filter and polarization-maintaining single-mode fiber, with a coupling efficiency of 75\%, are then exploited to detect the vertically polarized Stokes and anti-Stokes photons. Finally, the time-resolved coincidence counts of the biphotons are registered by two single-photon counting modules (60\% quantum efficiency) and a time-to-digital analyzer (1-ns time bin) for analyzing the temporal wave packet. This is done periodically to create a duty cycle 10\%, with atom cooling and trapping processes occurring for 4.5 ms followed by an experimental window of 500 $\mu$s. The trapping magnetic field is switched off during this experimental window to minimize the decoherence.

\begin{figure}
\includegraphics[width=8.5cm,height=8.5cm]{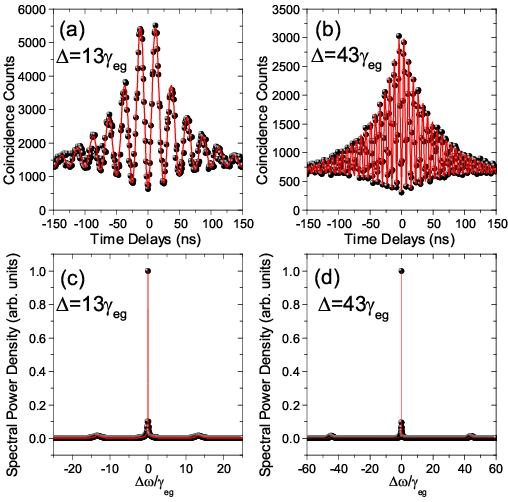}

\caption{\label{fig:3}The wave packets and the normalized cross-correlation functions $g^{(2)}(\tau)$ (insets) of the multi-mode biphotons are shown in (a) and (b) for $\vartriangle = 13\gamma_{eg}$ and $43\gamma_{eg}$, respectively, where $\varOmega_{p}=20$ MHz.
The spectral power densities of the biphotons in (a) and (b) are shown in (c) and (d), respectively. The dots are the experimental data points and the curves are the theoretical fits. The measurement times are 300 s and the time bin is 0.2 ns.}
\end{figure}

As depicted in Fig.~\ref{fig:2}(c), there are two sFWM resonant transitions that contribute to the generated Stokes and anti-Stokes fields. For the two-photon resonance shown on the left of Fig.~\ref{fig:2}(c), the generated Stokes and anti-Stokes fields have the same central frequencies as the pump beam. For the three-photon resonance shown on the right of Fig.~\ref{fig:2}(c), the resonances occur around $\omega_{p}$ and two virtual states are created due to the pump field, one at the frequency $\vartriangle$ above (blue-detuned) the ground state $\left|g\right\rangle $ and the other at the frequency $\vartriangle$ below (red-detuned) the excited state $\left|e\right\rangle $. The inelastically scattered pump beam then generates paired photons with central frequencies at $\omega_p\ \pm \vartriangle$. Neglecting propagation effect, the biphotons generated from both sFWM are approximately described by the following wave function \cite{Du:08, 35},
\begin{equation}
\Psi(t_{as},t_{s})\simeq-\frac{i\sqrt{\varpi_{as}\varpi_{s}}E^2_{p0}L}{\sqrt{8\pi}c}\int d\omega_{as}\chi_{as}^{(3)}e^{-i\omega_{as}\tau},\label{eq:2}
\end{equation}
where $L$ is the length of the atomic cloud, $\tau=t_{as}-t_{s}$ is the time delay between the detection of the anti-Stokes and Stokes photons, $\varpi_{as}(\varpi_{s})$ is the center frequency of the anti-Stokes (Stokes) photons, and $E_{p0}$ is the amplitude of the pump field. The third-order nonlinear susceptibility $\chi_{as}^{(3)}$ has a triplet structure corresponding to the two- and three- photon resonances. The linewidths of the central peak and the two sidebands are $\Gamma_{0} = \frac{\vartriangle^{2}}{\varOmega_{e}^{2}}\gamma_{g}+\frac{\left|\varOmega_{p}\right|^{2}}{\varOmega_{e}^{2}}\gamma_{eg}$ and $\Gamma  =  \frac{\left|\varOmega_{p}\right|^{2}}{\varOmega_{e}^{2}}\frac{(\gamma_{e}+\gamma_{eg})}{2}+\frac{\vartriangle^{2}}{\varOmega_{e}^{2}}\gamma_{eg}$, respectively, where $\gamma_{e}$ is the relaxation rate of the excited-state population, $\gamma_{g}$ is the dephasing rate of the inhomogeneously broadened ground state, $\gamma_{eg}$ is the dipole dephasing rate of the transition $\left|g\right\rangle \leftrightarrow \left|e\right\rangle$, $\varOmega_{p}$ is the pump's Rabi frequency, and $\varOmega_{e}=(\Delta^{2}+\left|\varOmega_{p}\right|^{2})^{1/2}$ is the effective Rabi frequency. In the presence of a weak pump or $\left|\varOmega_{p}\right| \ll \vartriangle$, $\Gamma_{0}$ approaches $\gamma_{g}$ and $\Gamma$ goes to $\gamma_{eg}=2\pi\times3$ MHz. The realization of sub-MHz-linewidth biphotons thus relies on both manipulating $\chi_{as}^{(3)}$ and suppressing the decoherence such as the temperature of the atoms, linewidth of the pump laser, and gradients of the earth and trapping magnetic fields.

Figure \ref{fig:3}(a) shows the measured biphoton wave packet with $\vartriangle=13\gamma_{eg}$, where the quantum interference between the wave functions of the two sFWM can clearly be seen. The wave packet can be described by the Glauber correlation function \cite{35, 37}, 
\begin{equation}
G^{(2)}(\tau) \propto [e^{-2\gamma_{e}\tau}+e^{-2\gamma_{g}\tau}-2{\rm cos}(\varOmega_{e}\tau)e^{-(\gamma_{e}+\gamma_{g})\tau}].
\label{eq:4}
\end{equation}
The first and second terms correspond to the sidebands and central component of $\chi_{as}^{(3)}$, respectively, and exhibit an exponential decay with the width inversely proportional to the linewidth. The third term represents the interference between the first two terms. The wave packet is thus a damped oscillation in which the oscillation period is determined by the effective Rabi frequency $\varOmega_{e}$ and the damping rate is determined by the linewidth $\Gamma_{0}$ of the inhomogeneously broadened ground state and the dipole dephasing rate $\Gamma$. 

\begin{figure}
\includegraphics[width=8.5cm,height=8.5cm]{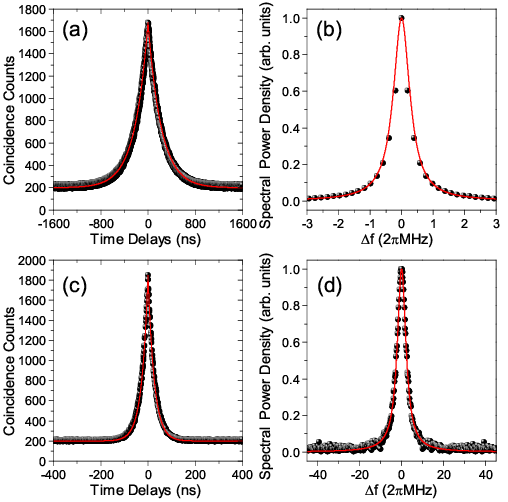}

\caption{\label{fig:5}(a) Wave packet and (b) spectral power density of the sub-MHz-linewidth biphotons. (c) Wave packet and (d) spectral power density of the biphotons associated with one sideband. The dots and curves are the experimental data points and theoretical fits, respectively. The error bars representing the statistical errors are similar to the size of the symbol.}
\end{figure}

 To generate sub-MHz-linewidth biphotons, we first increase the splitting of the central peak and sidebands in $\chi_{as}^{(3)}$ by increasing the pump detuning from $13\gamma_{eg}$ to $43\gamma_{eg}$. As shown in Figs.~\ref{fig:3}(c) ($\vartriangle=13\gamma_{eg}$) and \ref{fig:3}(d) ($\vartriangle=43\gamma_{eg}$), the resulting triplet are now well separated. Due to the larger detuning, the elastic scattering of the pump field is also reduced. The wave packet in Fig.~\ref{fig:3}(b) exhibits beating as before, but the oscillation period is larger because of the larger $\varOmega_{e}$. We next ``distill" the biphotons associated with the central peak by inserting a Fabry-Perot etalon with a bandwidth of 15 MHz and 60 MHz, respectively, in the anti-Stokes and Stokes channels. As shown in Fig.~\ref{fig:5}(a), the beating disappears and the wave packet has a double exponential waveform with a FWHM width of $\tau_{g}=513$ ns. The corresponding spectral power density is also shown in Fig.~\ref{fig:5}(b), which has an FWHM width of $0.11\gamma_{eg}=2\pi\times350$~kHz close to $\gamma_{g}=2\pi\times365$~kHz measured in the EIT measurement. The pair rate is measured to be 2,238 ${\rm s}^{-1}$ with a pump power of 500~$\mu$W. Correcting for the 60\% quantum efficiency of each single-photon detector, the 2.6\% (30\%) transmittance of the anti-Stokes (Stokes) channel including the transmission of etalon, and the duty cycle 10\%, the generation rate in the atomic ensemble is $7.97\times10^6$~${\rm s}^{-1}$. The spectral brightness is $2.28\times10^7$~${\rm s}^{-1}{\rm mW}^{-1}{\rm MHz}^{-1}$, which is 6 times higher than previously reported in a four-level atomic ensemble~\cite{Zhao2014} while the OD in our work is one order of magnitude lower. For comparison, we also distill the biphotons associated with the sidebands in $\chi_{as}^{(3)}$ by thermally tuning the etalons. The resulting wave packet and the spectral power density are shown in Figs.~\ref{fig:5}(c) and \ref{fig:5}(d), respectively. The temporal width is shorter as a result of the broader linewidth, which is limited to $\gamma_{eg}=2\pi\times3$~MHz. Moreover, because the wave packet is temporally symmetric, it suggests no preferred timing order of the detected Stokes and anti-Stokes photons--an interesting fact that is also evident by the equal transition probabilities in the dressed state picture. 

\begin{figure}
\includegraphics[width=8.5cm,height=8.5cm]{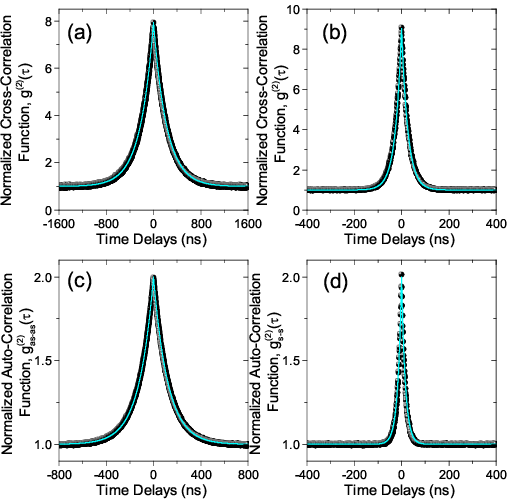}
\caption{\label{fig:6} The normalized cross-correlation functions of the (a) sub-MHz-linewidth biphotons and (b) biphotons associated with the sidebands. The normalized autocorrelation function of the (a) anti-Stokes photons, $g_{as,as}^{(2)}(\tau)$, and (b) Stokes photons, $g_{s,s}^{(2)}(\tau)$, as functions of the delay time.}
\end{figure}

The nonclassical correlation of the biphotons is verified by the Cauchy-Schwarz inequality $C=[g^{(2)}(\tau)]^{2}/[g_{as,as}^{(2)}(0)g_{s,s}^{(2)}(0)] \leq 1$, where $g^{(2)}(\tau)$ is the normalized cross-correlation function, $\tau$ is the time delay between the detection of the Stokes and anti-Stokes photons, and $g_{as,as}^{(2)}(\tau)$ and $g_{s,s}^{(2)}(\tau)$ denote the normalized autocorrelation functions of the anti-Stokes and Stokes photons, respectively. Figure \ref{fig:6}(a) shows the measured $g^{(2)}(\tau)$ of the sub-MHz-linewidth biphotons with a maximum value of $7.94\pm0.11$.The measured $g_{as,as}^{(2)}(\tau)$ and $g_{s,s}^{(2)}(\tau)$ are also shown in Figs.~\ref{fig:6}(c) and (d). At $\tau=0$, $g_{as,as}^{(2)}(0)=2\pm0.02$ and $g_{s,s}^{(2)}(0)=1.99\pm0.01$, inferring that the Stokes and anti-Stokes photons generated from spontaneous scattering are mostly thermal light. At very large $\tau$, $g_{as,as}^{(2)}(\tau)$ and $g_{s,s}^{(2)}(\tau)$ asymptotically approach 1, indicating that the photons detected by two detectors become uncorrelated. The Cauchy-Schwarz inequality is therefore violated with a maximum value of $C_{\rm max}=15.84\pm0.12$. For the biphotons associated with the sidebands, of which the measured $g^{(2)}(\tau)$ is shown in Fig.~\ref{fig:6}(b), the Cauchy-Schwartz inequality is violated with a peak value of $C_{\rm max}=20.71\pm0.16$. The violations thus confirm the nonclassical correlation between the Stokes and anti-Stokes photons for both types of narrowband biphotons.

\begin{figure}[t]
\includegraphics[width=8cm,height=5cm]{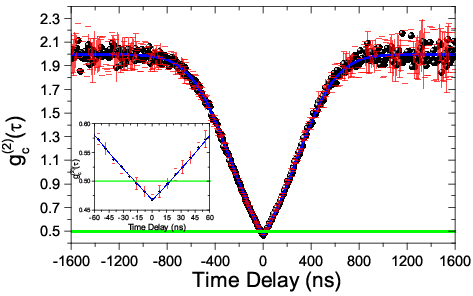}
\caption{\label{fig:7}Conditional auto-correlation of the heralded single photons as a function of the time delay \ensuremath{\tau}. The dots and curves are the experimental data points and theoretical fits, respectively. The time bin is 8 ns and the measurement time is 3,600 s. The green line indicates $g_{c}^{(2)}(\tau)=0.5$. }
\end{figure}

The strong non-classical correlation of the biphotons allows us to herald a single photon upon the detection of an anti-Stokes photon. To verify the single-photon nature of the heralded photons, the Hanbury-Brown-Twiss experiment is performed to measure the conditional autocorrelation function $g_{c}^{(2)}(\tau)$, which only exhibits the single-photon nature or $g_{c}^{(2)}(0)<0.5$ if $g_{s,as}^{(2)}(\tau) > 7.46$~\citep{39}. Experimentally, the Stokes photons are split into two channels by a 50:50 fiber-based beam splitter and the three-fold coincidences are detected between the anti-Stokes channel and the two split Stokes channels as shown in Fig.~\ref{fig:2} (a).  
In Fig.~\ref{fig:7}, the three-fold coincidences are measured as a function of the time delay. With $g_{c}^{(2)}(0)=0.46\pm0.01$, the antibunching is clearly observed and the single-photon nature of the heralded photons is confirmed. 

In conclusion, we have observed sub-MHz-linewidth biphotons in a two-level atomic ensemble by manipulating the biphoton wave functions of the two sFWM processes. The biphoton bandwidth is only limited by the ground-state decoherence. The temporally long biphotons also allow for the arbitrary shaping of the wave function and the generation of Bell states with a sub-MHz linewidth. The feasibility of generating heralded single photons in a two-level atomic ensemble is useful for diverse quantum applications including LOQC and quantum key distribution. Realization of a narrower bandwidth is possible if the residual magnetic gradient can be compensated by using additional coils. Achieving a higher pair rate or brightness is also possible by increasing the etalon's bandwidth in the anti-Stokes channel or increasing the duty cycle with the implementation of double MOTs, dark MOT, or faster switching circuit for the magnetic field. 

This work was supported by the National Science and Technology Council, Taiwan (Grant No. 113-2119-M-007-012).

\nocite{*}
\bibliography{REF}

\end{document}